François Renaville and Fabienne Prosmans

# Using LibCal Seats to Better Serve Students

**Abstract:**

This chapter examines the evolution of library services at the University of Liège (ULiège), with a focus on the implementation and assessment of the LibCal Seats booking module. Introduced in September 2020 in response to the COVID-19 pandemic, this system was designed to manage occupancy and maintain social distancing. While initially a temporary measure, the seat booking service remains in use during peak periods. Drawing on survey data from 2022 and 2023, the chapter analyses user perceptions of the system. Results indicate strong student appreciation, particularly regarding stress reduction and equitable access to study spaces. Despite overall satisfaction, issues such as unoccupied reserved seats and an unnecessarily complex booking process emerged, leading to targeted improvements. This chapter highlights the importance of responsive, user-centred services in academic libraries. The adoption of the booking system helped address challenges such as overcrowding and "seat hogging," ultimately contributing to a more organised and accessible environment. The case study illustrates how technology can enhance library service delivery, offering insights for institutions seeking to optimise space management. The continued evaluation of the system reflects a broader commitment to adapting services in alignment with user needs and institutional priorities.

## Introduction

The University of Liège (ULiège) is a medium-sized French-speaking university founded in 1817 and originally located in the city centre of Liège, Belgium. In 1954, the decision was taken to partially move the university to the suburb of Sart Tilman, located approximately 10 km from the city centre. In the 2000s and 2010s, the university merged with other Belgian research and teaching colleges. ULiège now has four campuses across three cities, namely Liège, Gembloux and Arlon. Currently, ULiège is composed of eleven faculties, with more than 26,600 registered students and 5,700 faculty and staff members (University of Liège 2024).

  The relocation of ULiège to the Sart Tilman campus in 1956 led to a radical transformation of its libraries and information services, marked by the creation of specialised *unités de documentation* (documentation units) that were close to users and, therefore, related to faculties or departments. All these units were characterised by a high degree of autonomy and independence. In the early 2000s, ULiège modernised its libraries and documentation units, reorganised them extensively, and grouped them into five major entities. Several units and libraries merged (Cuvelier, Vanhoorne, and Thirion 2008). At the end of the 2010s, ULiège approved the strategic reorientation of library management, moving towards a new model that centred on services, termed "Library as a Service".

The ULiège Library comprises fifteen distinct library branches spread over the four campuses and offers a total of 2,222 seats, with a provision of fewer than one place for ten students. The total area of the ULiège Library spaces accessible to users is 12,385 m$^2$, or 0.48 m$^2$ per user. This figure is significantly lower than the average for Belgian universities (Comité de direction et responsables de proximité de ULiège Library 2023). The demand for library seats is, therefore, particularly high during periods of study cramming and exam periods.

**Library Seat Booking Services**

While the ULiège Library has offered a collaborative working room reservation service since 2013, a seat booking service was not available until September 2020. A first-come, first-served rule was applied. As in many other institutions and companies, the COVID-19 pandemic acted as a game changer. In September 2020, at the start of the new academic year, the library responded to COVID-19 public health regulations by deploying a seat-booking service in five branches to avoid queuing, unnecessary trips to the library, and to enable users to spread out in the library spaces to respect social distancing and the maximum occupancy rate (Wiley and Gowing 2020). At that time, the duration of the booking periods was limited to four consecutive hours. Students were asked to leave the premises during two half-hour periods for ventilation purposes, specifically at 12:30 and 16:00.

The LibCal Seats module was employed for the booking process. Known to users as Book It (ULiège Library n.d.b), the LibCal solution has been in use since 2019 to manage the reservation of group study rooms (LibCal Spaces) and the promotion of library events and library workshops (LibCal Events). LibCal Seats is a complementary module to LibCal Spaces. The module was released by Springshare in July 2020, during the pandemic, to meet libraries' needs to manage their attendance (Talia 2020b). The ULiège Library is one of the first European libraries that subscribed to LibCal Seats.

Since the September 2021 academic year, when the COVID-19 pandemic began to decline, the obligation to reserve seats was no longer enforced. However, the service was kept in place in two branch libraries during rush hours and exam periods. LibCal Seats was reused for two locations, the Reading Room and *Santé-CHU* (the Health Library), between December 11, 2021 and January 28, 2022, and between May 21, 2022 and June 24, 2022, to offer "equitable and maximised library access for students" (Wiley and Gowing 2020). The Reading Room is located in a building adjoining the Faculty of Philosophy and Letters on the 20-Août campus; the Health Library is on the Sart Tilman campus (suburb) and primarily covers disciplines related to the Faculties of Medicine; Veterinary Medicine; and Psychology, Speech Therapy, and Education Sciences.

From December 2021, the ULiège Library has activated the check-in and check-out options in LibCal. Once a booking is made via an online form, where the student selects the specific seat, date, and time slot they wish to reserve, they receive an alphanumeric code. This code must be entered on a web page accessible via a QR code upon arrival to confirm their check-in. If there is no check-in, the booking is automatically cancelled within thirty minutes of the start of the booking. This measure resolves the problem of unoccupied reserved seats (Breen, Dundon, and McCaffrey 2018). If users leave earlier than expected, they are asked to check out online by reusing the same alphanumeric code used for check-in, thereby freeing up the remaining time for others.

After two years of experience with LibCal Seats, the ULiège Library decided to conduct a survey to collect feedback and analyse data to improve the booking service.

**Libraries and Seat-Booking Services**

Numerous studies have been conducted to understand how and which library spaces and furnishings are used, what users like and dislike about them, and why they avoid or prefer specific spaces or floors (Applegate 2009; Attebury, Doney, and Perret 2020; Cha and Kim 2020; Jaskowiak et al. 2019; İmamoğlu and Gürel 2016; McGinnis and Kinder 2021). As Nichols and Philbin (2022) demonstrated, an appropriate balance between seating quality and quantity is necessary. The impacts of environmental factors such as printers, food services, exhibits, art displays, washrooms, and walkways through study spaces within the library should be considered by librarians and space designers. Additionally, the perception of the occupancy rate of a library space is subjective. A number of studies confirm that spaces are perceived to be "full" when only half of the seats per table are occupied (Khoo et al. 2016, 66; Gibbons and Foster 2007, 28).

Nelson (2016) extensively studied the LibCal product; however, as the solution is continually improved and new functions are added over time, Nelson's review is dated in some respects. In a more recent study, Scronce (2023) described the implementation and use of the LibCal Space module to manage room and space reservations at the College of Charleston. As the LibCal Seat module is a new feature of the LibCal suite (Talia 2020a, 2020b), the literature on the subject is sparse.

A number of libraries introduced seat-booking services during the COVID-19 pandemic. Wiley and Gowing (2020) presented their study on the successful utilisation of LibCal to allow seat-level appointments to limit library access during the pandemic at the University of Miami Libraries. ULiège has had a similar experience and uses the system in a post-pandemic context as well. The University of Florence's *Biblioteca di Scienze Sociali* had a less promising experience, where the implementation of seat reservation was strongly criticised by students and inspired some of the most incisive expressions of indignation and discontent (Melani and Paoletti 2022).

Students continue to believe that a seat reservation service is needed in the post-pandemic period. Atkinson (2021) stated, "[a]s a result, many librarians [think] that they [are] likely to retain seat booking systems introduced during the pandemic, either as a permanent feature for a percentage of study and PC spaces or at particular times of the year such as examination periods" (307). Wales (2021) also identified that there is a requirement for selected seating areas and not necessarily for the whole library.

There are real opportunities for libraries to improve user services by setting up a more systemic reservation system. Students queuing outside the university library building in China have been reported in the media (Ye 2017), proving that seat requests can be extremely high at peak periods. Seat booking services offer a way to limit the "towels on deck chairs" phenomenon (Breen, Dundon, and McCaffrey 2018, 105), also called "seat hogging" (Ishak and Ong 2016), by students reserving library seats for themselves because they are afraid of not having a seat or for friends scheduled to arrive later (Lau et al. 2015). Desk-clearing interventions by library staff can reduce unfair behaviour (Breen, Dundon, and McCaffrey 2018), similar to an integrated control system of check-in and check-out considered in this study, or a more elaborate system integration using seat pressure sensors (Welsen 2022).

**Satisfaction Surveys**

On June 20, 2022, 2,269 ULiège students and university members who had made at least one reservation for a seat in the Reading Room or Health Library since May 21, 2022, received the first anonymous satisfaction survey by email. The survey comprised eighteen questions focusing on users' habits and their experience with the booking service. It closed on July 1, 2022, and received 375 complete responses, representing a response rate of 16.5%.

Actions have been taken and measures implemented to improve the service based on the analysis of these results (Renaville 2022). Changes were implemented in 2023. To measure the satisfaction rate and ensure that students appreciated the improvements, a follow-up survey was carried out between October 4, 2023, and October 31, 2023.

Of the 6,737 students contacted, all of whom made at least one seat booking at the Reading Room or Health Library between May 2023 and June 2023, 315 students responded to the survey (rate 4.7%; Renaville and Prosmans 2023). Most final-year master's students probably did not respond to the survey, as it is likely that they did not check their email inbox or feel as concerned about the changes as other students. The second survey was also anonymous. Comprising only ten questions, it was much shorter, and it focused only on measuring the satisfaction rate after the implementation of the changes. Moreover, the survey contained no questions regarding faculty affiliation, gender or attendance visits.

**An Appreciated Solution**

Of the 375 respondents to the first survey, 255 were enrolled in bachelor's programmes and 120 in master's programmes. Their faculty affiliation is presented in Table 5.1: Student Respondents by Faculty.

| Faculty | # |
|---|---|
| Applied Sciences | 43 |
| Architecture | 5 |
| Gembloux Agro-Bio Tech | 1 |
| HEC Liège School of Management | 32 |
| Law, Political Science, and Criminology | 34 |
| Medicine | 117 |
| Philosophy and Letters | 31 |
| Psychology, Speech Therapy, and Education Sciences | 28 |
| Sciences | 26 |
| Social Sciences | 10 |
| Veterinary Medicine | 46 |
| Other | 2 |
| **Total** | **375** |

Table5.1: Student respondents by faculty in the 2022 survey.

Regarding the primary user group of the two library branches where seat booking was enabled in 2022, the Health Library, including students from the Faculty of Sciences (biology, biochemistry, and others) accounts for 191 respondents. For the Reading Room, which is located near the Faculty of Philosophy and Letters on the 20-Août campus, only 31 respondents are from the humanities. The answers to the questions on affiliation and visits show that the use of these two spaces goes well beyond disciplinary divisions, and students also frequent libraries that are not within their discipline.

Students were first asked which libraries they frequented most often from amongst the library branches that require seat bookings. Around half of the respondents (48.3%) used the Health Library exclusively. The Reading Room was used exclusively by a third (35.5%) of the respondents. A small proportion (16.3%) visited both sites.

Of the 133 respondents who visited the Reading Room exclusively, twenty-seven (20.3%) came from the Faculties of Medicine (10); Veterinary Medicine (3); and Psychology, Speech Therapy, and Education Sciences (14). Furthermore, twenty-eight (21.1%) respondents belonged to the Faculty of Philosophy and Letters. HEC Liège School of Management and the Faculty of Applied Sciences accounted for twenty-two (16.5%) and twenty (15.0%) of the respondents, respectively.

Of the twenty-seven respondents who mainly visited the Reading Room but sometimes also went to the Health Library, twenty-one came from the faculties of Medicine (14); Veterinary Medicine (3); and Psychology, Speech Therapy, and Education Sciences (4).

Of the 181 respondents who visited the Health Library exclusively, 124 (68.5%) were from the faculties of Medicine (80); Veterinary Medicine (37); and Psychology, Speech Therapy, and Education Sciences (7). Sixteen (8.8%) respondents were from the Faculty of Applied Sciences and fifteen (8.3%) were from the Faculty of Law, Political Science, and Criminology.

Of the twenty-three respondents who mainly visited the Health Library but sometimes also went to the Reading Room, sixteen were from the Faculties of Medicine (10); Veterinary Medicine (3); and Psychology, Speech Therapy, and Education Sciences (3).

The students were asked if they found it practical to be required to book their place for these two branches during peak periods. A significant majority of respondents (81.6%, 306) found it practical, while 17.6% (66) were against the service.

There was no significant difference between bachelor's and master's students' opinions. Students from the Faculty of Veterinary Medicine constituted the most supporters of the booking obligation (89.1%, 41 out of 46). The second-highest number of students were in the Faculty of Medicine (83.8%, 98 out of 117); however, they represented twice as many responses in absolute terms.

**Compulsory Booking Service**

Students were asked if they would have liked to see a compulsory booking service introduced in other library branches during peak periods. The responses were fairly evenly balanced. One hundred and twenty (32.0%) students were in favour of a compulsory booking service, 118 (31.5%) students were against it, and 137 (36.5%) of them had no opinion. Of the 120 people who answered "yes", 103 (85.8%) students mentioned the Léon Graulich library branch, nineteen (15.8%) of them named the Sciences branch, fifteen (12.5%) the Germanic Languages and Literatures branch, thirteen (10.8%) mentioned the Historical Sciences site, and eleven (9.2%) named the Romance Languages and Literatures branch (multiple answers were possible). Léon Graulich, one of the major libraries with law, management, and social sciences collections, and the Sciences branch were two library branches where LibCal Seats had been enabled in 2020 at the very beginning of the pandemic.

Of the 103 students who mentioned the Léon Graulich site, thirteen (12.6%) of them were from the Faculty of Law, Political Science, and Criminology; ten (9.7%) were from HEC Liège School of Management School; four (3.9%) were from Social Sciences; twenty-four (23.3%) were from the Faculties of Medicine; fifteen (14.6%) were from Veterinary Medicine; fourteen (13.6%) were from Psychology, Speech Therapy, and Education Sciences; and twelve (11.7%) were from Applied Sciences. The remaining eleven were spread across the other faculties.

## Student suggestions

### Booking at Other Branches

Nearly 60 people commented on why it would be useful if seats could also be booked in other library branches which don't offer this service. The most frequently cited reasons are as follows:
- It ensures finding a place to work, which saves time and avoids unnecessary trips to the library (Lau et al. 2015).
- It reduces stress.
- It reduces conflict, particularly if users 'reserve' places for friends who may arrive much later (Breen, Dundon, and McCaffrey 2018).
- It prevents students who are not from the University from accessing the library to work.

For this reason — namely, to prevent access by non-University students —, the smallest branches on the city centre campus were often cited.

In August 2022, before obtaining the results of the survey, the ULiège Library decided to enable the seat booking service at the Léon Graulich branch beginning December 2022. This step was taken anticipating an extremely strong demand formulated by many survey respondents. The 2023 follow-up survey showed that 73.0% of the respondents (230) appreciated that it had become possible to book a seat at Léon Graulich too.

In December 2023, due to high demand and the risk of saturation of the smallest spaces on the city centre campus, seat booking was also enabled in three additional branches located in the same facilities as the Faculty of Philosophy and Letters, namely Historical Sciences, Germanic Languages and Literatures, and Romance Languages and Literatures. In addition to the 335 existing seats in the Reading Room, 226 new seats for these three branches were made available in Book It on the 20-Août campus.

### Cohesive and Common Rules

Students were also asked about the uniformity of reservation rules and whether they would prefer identical booking rules across all branches. More than half of the respondents (51.2%) preferred uniform rules. Furthermore, 125 (33.3%) students stated that they did not mind if the rules varied from branch to branch (for example, the maximum booking length or the maximum number of bookings allowed could differ from one site to another). Lastly, 58 (15.5%) students had no opinion.

**Booking at Other Times than Peak Periods**

A significant majority (279, 74.4%) of students expressed disapproval regarding whether compulsory seat booking would merit being introduced during times other than the busy periods of peak studying and exam periods. Only 51 (13.6%) students were in favour of an extension of the service, such as, for example, just before the beginning of the cramming period.

**Confirmation Email**

When questioned regarding the clarity of the booking confirmation email, 352 (93.9%) students answered that the email was clear or fairly clear. Among the few comments and suggestions for improvement, there was a repeated request relating to a more visible and direct display of the validation code for the check-in as noted in the following section.

**Reminder Email**

The library had set up an automatic reminder email to be sent twenty-four hours before the start of the booking. Of the 375 respondents, 259 (69.1%) students stated that receiving a reminder was useful or fairly useful, and 97 (25.9%) of them thought that the reminder email was useless or fairly useless. The most common explanations given were as follows:

- Seat booking is generally done a few days in advance. Students also indicate in their agenda when they are going to work in the library. As the electronic calendar can generate a reminder, there can be quite a few reminders for the same booking, particularly if there is more than one booking on the same day.
- Students feel they already receive a lot of emails, and their email inboxes are overloaded.
- It is not environmentally friendly.

Even though most respondents were in favour of the reminder in the 2022 survey, the library decided to suspend this service because it was sensitive to the arguments and comments of those who found it not helpful. In the 2023 survey, 169 respondents (53.7%) found this deactivation positive, and 60 (19.0%) of them regretted it.

**Booking Validation**

When asked how much they liked having to confirm their presence through a code within thirty minutes of the start of the reservation, 198 respondents (52.8%) found it rather practical, 123 (32.8%) found it rather restrictive, and 34 (9.1%) found it unnecessarily restrictive. Twenty (5.3%) students had no opinion on the matter.

In total, 157 respondents found validation rather restrictive or unnecessarily restrictive in the workflow. Of these, 141 students reported that it was because they sometimes forget to validate their presence, 33 admitted to being unable to find the validation code, and thirteen considered the process complicated (multiple answers were possible). Several respondents also indicated that the 30-minute time limit granted by the system to validate a reservation was too short. The most common reasons they cited included an exam taking longer than expected, and traffic jams or delays in public transport.

In 2022, LibCal offered a maximum of thirty minutes to validate a reservation. Since the June 2023 release, however, it is possible to select 60 minutes or 120 minutes to make a confirmation. These check-in delay options can be significantly longer during busy periods, and the library has decided to retain the 30-minute limit.

Despite the restrictive aspect of validation, several respondents acknowledged that this measure allows for freeing up seats booked by students who do not come to the library in the end. The validation system does not prevent students from validating their bookings remotely, that is, without being physically present in the library. Some students have understood this rule, and they validate their booking on time, but well before arriving at the library. The 2023 follow-up survey confirmed this finding, and many respondents deeply regretted that it was possible to check in remotely. This weakness in the LibCal system was pointed out to Springshare, which developed a geolocation-based check-in option in 2024 to prevent remote check-ins (Reed 2024).

Notably, until the May to June 2023 exam session, it was possible to book a seat up to ten days in advance. Since December 2023, this limit has been reduced to three days based on other libraries' experiences (SciencesPo Bibliothèque 2023). The intended effect is to limit the practice of chain booking, that is, the repeated reservation of library study spaces by the same individuals or groups over consecutive days, often without certainty of actual use. This practice was frequently criticized in the 2023 survey by some students who, despite their best intentions, could not reliably predict their availability to attend the library five to ten days in advance. With a limit of three days, a reminder is even less necessary. Some libraries, such as the University of Miami's library, have configured a window of one day for advance bookings (Wiley and Gowing 2020).

**Ease of Use**

When asked about the ease of use of the booking service, 324 (86.4%) students thought it was easy or fairly easy to book a seat in the library. More than 35 people explained why they found it difficult or rather difficult to book a seat online. The following comments were made most frequently:

- Users must book early enough. A slightly last-minute booking sometimes has no chance of succeeding as all seats are already booked.
- It is difficult to visualise the seats in the library and recalling where exactly they are located in the environment (for instance, next to windows, power sockets, doors, lifts, busy passages, or washrooms (Nichols and Philbin 2022).
- Some students admit that they forget to make a reservation before coming to the Health Library or Reading Room (Melani and Paoletti 2022).
- The booking service website on mobile devices has poor usability, and it is much easier to make a booking from a computer.
- It is difficult to find the validation code to confirm attendance once a student arrives at the library (that is, to check in) because too many emails are sent for various bookings, which sometimes makes it difficult to find the right code. This issue is compounded by the limited usability of the Zimbra Collaboration Suite (ZCS) webmail client on smartphones that is used at ULiège.

**Unwanted Occupancy**

The ULiège Library was interested in determining the extent to which students felt that their peers respected the reservation system and workflow. When asked if the seat they had

Renaville, F. & Prosmans, F. (2026). Using LibCal Seats to better serve students. In C. Furno, M. K. Saba, and M. Stöpel (Eds.), *Changing information services and user experiences*. De Gruyter Saur.
https://hdl.handle.net/2268/338194

booked was occupied by someone else, 278 (74.1%) of the students said yes, while 92 (24.5%) of them stated that they had never experienced such a mishap. The fact that almost three-quarters of respondents experienced this problem is highly surprising. Several respondents considered their overall experience with Book It (since 2020) rather than the reference month (between May 21 and June 24, 2022), which partly explains the high rate of positive responses. If this were the case, these figures need to be put into perspective.

Finally, 241 respondents (64.3%) admitted to having already occupied a seat that they had not reserved (because they forgot to make a booking or because their reserved seat was occupied by another student). Such inappropriate behaviours and instances of unwanted occupancy contribute to a cumulative chain reaction of negative experiences, which undermines efforts to maintain a calm and respectful atmosphere in the library. Students should be made more aware of this issue.

### Actions Taken and Follow-up

To consider these comments and improve the situation, the ULiège Library made several decisions effective for the May/June 2023 booking period. Their effectiveness was measured by the October 2023 follow-up survey. These implementations and further recommendations are given below:

- Increase in on-site posters: While posters and roll-ups had been deployed at the Reading Room and Health Library, less than a third of respondents to the Autumn 2023 survey appeared to have noticed an improvement. From the comments submitted, it seems - hat the library needs to continue making an effort to improve communication.
- Improvement in the visibility of library plans on the booking platform and in emails: Despite the changes made, it seems that there is a need for a continued effort to improve the visibility of library plans. The ULiège Library also decided to subscribe to an additional service at Springshare, namely interactive maps, which could go a long way in improving the situation. Unfortunately, it was not possible to deploy the new module in 2023.
- Compatibility on mobile devices: LibCal's poor compatibility with mobile devices is a well-known weakness of the system, and the customer community is waiting for a significant improvement. This change particularly concerns the availability grid for selecting the time slots and navigating through them. Springshare is aware of this weakness and intends to work on it. In its June 2023 release, Springshare deployed Search by Space, a new search form based on the recommendations of available seats and rooms (Kyoya 2023). While some respondents to the follow-up survey were unconvinced, the 2023–2024 winter cramming period was an opportunity to fully test the new form.
- More visible check-in validation code: The check-in validation code has been made more visible in the booking confirmation email. According to the follow-up survey's results, 284 of the 315 respondents (90.2%) appreciated this change. From December 2023, the code is displayed in the subject line of the email for several library branches.

Besides these decisions taken by the library after the 2022 survey, the technical team comprehensively reviewed the public interface of Book It (ULiège Library n.d.a) by integrating it more into the graphic charter of the library website, making it more responsive and offering a multilingual experience (ULiège Library n.d.b). The 2023 follow-up survey indicated that 144 respondents (45.7%) found the new layout better and more useful to navigate, while only ten (3.2%) respondents did not.

A four-hour booking slot limit was implemented in 2020–2021 and during the winter of 2021–2022, particularly because of the need to ventilate the sites frequently during COVID-19. Every four hours or so, students were asked to leave, and the windows were opened. Since May 2022, the maximum total booking time for a seat has been increased from four to ten hours. If students wish to stay for longer than ten hours in a library, they must create a second reservation. Of the 375 respondents, 228 (60.8%) students found ten hours to be sufficient, 99 (26.4%) of them found it to be too short and would have preferred the option to book a slot for a whole day, and twelve (3.2%) students found it to be too long. The 2023 follow-up survey confirmed that the extension to ten hours was greatly appreciated by 288 (91.4%) of the 315 respondents.

A positive consequence of the extension to ten hours is the fact that it reduces the number of emails sent to users, which was frequently expressed in the students' feedback. However, the following points should be noted about the extension:
- The extension should encourage the library to make users more aware of the need to check out if they leave the library earlier than expected, so as not to block an available space.
- The extension aggravates the phenomenon of remote validation by students when it is followed by a no-show (see below).
- The extension can prove problematic in the pre-cramming period when courses are still being conducted. Some students book a seat for ten hours but attend courses during the day. In such cases, booked seats may be left unoccupied for a long time. Although a shorter maximum period could be enforced (such as a return to a four-hour booking limit), it could make things less clear for users who prefer uniform and coherent rules across all locations.

For an efficient and fair booking service, the library relies on students' cooperation and fair play; therefore, awareness campaigns were planned to avoid unoccupied seats as much as possible (Breen, Dundon, and McCaffrey 2018).

Finally, the fact that many students must search for the code in their email inbox seems to indicate that not all of them save the booking in their calendar or even use a calendar. An .ics calendar file is always sent as an attachment to each booking confirmation, and the code is prominently displayed in the event created in the agenda so that students can easily access the validation code in their calendars. Specific communication from the library would undoubtedly be beneficial—for instance, sending reminders about check-in procedures, providing tips for managing bookings effectively, or issuing targeted messages to users who frequently miss or duplicate reservations.

## Conclusion

A seat-booking service was initially set up in September 2020, during the COVID-19 pandemic, in five branches of the ULiège Library. This service was enforced to respect social distancing, manage the flow and avoid queues at the entrance of the library. The service was reactivated in two of the three main branches during the December 2021 to January 2022 and May 2022 to June 2022 peak periods. During the summer of 2022, a satisfaction survey was conducted among students who used seat booking, aiming to improve the service. To consider user expectations, several decisions were taken and implemented for the May to June 2023 booking period, including increasing on-site posters, improving the visibility of plans on Book It and in emails, enhancing the compatibility on mobile devices, improving the visibility of the validation code for check-in and check-out, and extending the booking period from four hours to a maximum ten of ten hours. One of the main expectations, namely the obligation to reserve a place in the third major branch, was anticipated and implemented during the cramming period between December 2022 and January 2023. To solicit the students' opinions on these changes, a second survey was conducted in autumn 2023.

Overall, the recommendations issued following the 2022 survey had benefits, even if some points still require improvement, such as continued communication efforts. The Book It service and the LibCal solution's application and configuration are not to everyone's satisfaction, as some students still prefer the previous first-come, first-served solution. However, it is fair to say that the maximum of the system's current configuration possibilities has been reached. The introduction of new functions, such as a geolocation-based check-in option — added by Springshare in March 2024, activated at ULiège in April 2024, and subsequently evaluated through a satisfaction survey conducted the same year — and a more responsive availability grid, addresses students' needs. For the next peak periods, the ULiège Library plans to continuously improve the seat booking service in line with user expectations.

## Acknowledgements

We would like to thank Robert De Groof for his invaluable technical expertise on the new version of Book It. We are also grateful to our colleague Tommy Deswysen for his suggestions.

## References

Applegate, Rachel. 2009. "The Library Is for Studying: Student Preferences for Study Space." *The Journal of Academic Librarianship* 35 (4): 341–346. https://doi.org/10.1016/j.acalib.2009.04.004.

Atkinson, Jeremy. 2021. "The Times They Are A-changin': But How Fundamentally and How Rapidly? Academic Library Services Post-pandemic." In *Libraries, Digital Information, and COVID: Practical Applications and Approaches to Challenge and Change*, edited by David Baker and Lucy Ellis, 303–315. Chandos Publishing. https://doi.org/10.1016/B978-0-323-88493-8.00019-7.

Attebury, Ramirose Ilene, Jylisa Doney, and Robert Perret. 2020. "Finding Our Happy Place: Assessing Patron Satisfaction after a Comprehensive Remodel." *Journal of Learning Spaces* 9 (2): 22–32. https://libjournal.uncg.edu/jls/article/view/1929.

Breen, Michelle, Mary Dundon, and Ciara McCaffrey. 2018. "Making Every Seat Count: Space Management at Peak Times in a University Library." *New Review of Academic Librarianship* 24 (1): 105–118. https://doi.org/10.1080/13614533.2017.1414066.

Cha, Seung Hyun, and Tae Wan Kim. 2020. "The Role of Space Attributes in Space-choice Behaviour and Satisfaction in an Academic Library." *Journal of Librarianship and Information Science* 52 (2): 399–409. https://doi.org/10.1177/0961000618794257.

Comité de direction et responsables de proximité de ULiège Library. 2023. "ULiège Library : Rapport annuel 2022.", edited by Stéphanie Simon and Paul Thirion. Liège: ULiège Library. https://hdl.handle.net/2268/306299.

Cuvelier, Catherine, Frédérick Vanhoorne, and Paul Thirion. 2008. "Le Réseau des Bibliothèques de l'ULg." *Lectures* 158, Novembre-Décembre: 18–25. https://hdl.handle.net/2268/642/.

Gibbons, Susan, and Nancy Fried Foster. 2007. "Library Design and Ethnography." In *Studying Students: The Undergraduate Research Project at the University of Rochester*, edited by Nancy Fried Foster and Susan Gibbons, 20–29. Chicago: ALA. http://hdl.handle.net/1802/7520.

İmamoğlu, Çağrı, and Meltem Ö. Gürel. 2016. "Good Fences Make Good Neighbors: Territorial Dividers Increase User Satisfaction and Efficiency in Library Study." *The Journal of Academic Librarianship* 42 (1): 65–73. https://doi.org/10.1016/j.acalib.2015.10.009.

Ishak, Yuyun W., and Vincent Ong. 2016. "Library and Experiential Learning: Taking Collaboration to the Next Level." *Proceedings of the IATUL Conferences*. Paper 1. https://docs.lib.purdue.edu/iatul/2016/infolit/1.

Jaskowiak, Megan, Kari Garman, Meg Frazier, and Todd Spires. 2019. "We're All in This Together: An Examination of Seating and Space Usage in a Renovated Academic Library." *Library Philosophy and Practice (e-journal)*. 2645. https://digitalcommons.unl.edu/libphilprac/2645.

Khoo, Michael J., Lily Rozaklis, Catherine Hall, and Diana Kusunoki. 2016. "'A Really Nice Spot': Evaluating Place, Space, and Technology in Academic Libraries." *College & Research Libraries* 77 (1): 51–70f. https://doi.org/10.5860/crl.77.1.51.

Kyoya. Juri. 2023. "Code Release: LibCal, LibAnswers, LibGuides, LibStaffer, LibInsight, & LibWizard Features Coming Your Way!" *The Springy Share*. June 23. https://blog.springshare.com/2023/06/23/code-release-libcal-libanswers-libguides-libstaffer-libinsight-amp-libwizard-features-coming-your-way/.

Lau, Aldred Wen Yang, Eileen Yi Lin Tan, Jo Xin Lee, et. al. 2015. "Improve Space and Manpower Utilisation." Singapore: Singapore Management University. https://ink.library.smu.edu.sg/library_research/55/.

McGinnis, Robbin, and Larry Sean Kinder. 2021. "The Library as a Liminal Space: Finding a Seat of One's Own." *The Journal of Academic Librarianship* 47 (1): 102263. https://doi.org/10.1016/j.acalib.2020.102263.

Melani, Chiara, and Alessio Paoletti. 2022. "Tu, il Covid e la biblioteca." *Biblioteche oggi* 40 (6): 53–63. http://dx.doi.org/10.3302/0392-8586-202206-053-1.

Nelson, Gregory M. 2016. "LibCal: A Product Review." *The Reference Librarian* 57 (1): 57–72. https://doi.org/10.1080/02763877.2015.1096226.

Nichols, Aaron, and Paul Philbin. 2022. "Library Usage Study, the How and What: A Survey of Space Usage at a Mid-Sized Research Library." *Evidence Based Library and Information Practice* 17 (4): 122–138. https://doi.org/10.18438/eblip30103.

Reed, Rebecca. 2024. "Code Release: LibCal, LibAnswers, LibStaffer, LibGuides, and LibInsight Features Coming Your Way!" *The Springy Share*. March 15.

https://blog.springshare.com/2024/03/15/code-release-libcal-libanswers-libstaffer-libguides-and-libinsight-features-coming-your-way/.

Renaville, François. 2022. "Enquête sur la réservation de place en bibliothèque – Analyse des résultats de l'enquête de juin 2022 menée à ULiège Library." Liège: ULiège Library. https://hdl.handle.net/2268/296315.

Renaville, François, and Fabienne Prosmans. 2023. "Enquête de suivi sur la réservation de place en bibliothèque – Analyse des résultats de l'enquête d'octobre 2023 menée à ULiège Library." Liège: ULiège Library. https://hdl.handle.net/2268/309367.

SciencesPo Bibliothèque. 2023. "Réservation de places – Enquête fev. 2023." Paris: SciencesPo Bibliothèque. https://www.sciencespo.fr/bibliotheque/sites/sciencespo.fr.bibliotheque/files/files/pdfs/Infographie_resultats_enquete_reservation_de_places_2023.pdf.

Scronce, Gretchen. 2023. "Streamlining Room Scheduling and Managing Library Events Using LibCal." *Journal of Web Librarianship* 17 (4): 75–94. https://doi.org/10.1080/19322909.2023.2263648.

Talia. 2020a. "LibCal Seats Module – Safely Reopen Your Building (Blog Series: Part One)." *Springshare Blog*, June 17. https://web.archive.org/web/20250115132714/https://blog.springshare.com/2020/06/17/libcal-seats-module-safely-reopen-your-building-blog-series-part-one/.

Talia. 2020b. "LibCal Seats Is Just Around the Corner!" *Springshare Blog*. July 13. https://web.archive.org/web/20250214151804/https://blog.springshare.com/2020/07/13/libcal-seats-is-just-around-the-corner/.

ULiège Library. n.d.a. "Book it, votre système de réservation." Accessed March 27, 2023. https://web.archive.org/web/20230327082700/https://uliege.libcal.com/.

ULiège Library. n.d.b. "Book it, votre système de réservation." Accessed April 26, 2025. https://bookit.lib.uliege.be.

University of Liège. 2024. "Key Figures." Last modified September 16. https://www.uliege.be/cms/c_11478577/en/key-figures.

Wales, Tim. 2021. "Back to the Future? Practical Consequences and Strategic Implications of a UK Academic Library's COVID Response." In *Libraries, Digital Information, and COVID: Practical Applications and Approaches to Challenge and Change*, edited by David Baker and Lucy Ellis, 21–29. Cambridge, MA: Chandos Publishing. https://doi.org/10.1016/B978-0-323-88493-8.00021-5.

Welsen, Sherif. 2022. "Design of an Innovative Campus Remote Seat Booking System for Smart Learning Environment." In *2022 4th International Conference on Computer Science and Technologies in Education (CSTE)*, 251–254. IEEE. https://doi.org/10.1109/CSTE55932.2022.00053.

Wiley, Glen, and Cheryl Gowing. 2020. "Starting Back: Using LibCal Seats to Manage Users and Spaces at the University of Miami Libraries." *Association of Southeastern Research Libraries (ASERL) Webinar Series*, August 21. https://scholarship.miami.edu/esploro/outputs/991031769621002976.

Ye, Josh. 2017. "Chinese University Students Queue for Hours for Library Study Space." *South China Morning Post*, January 5. http://www.scmp.com/news/china/society/article/2059620/chinese-university-students-queue-hours-library-study-space.